\documentclass[aps,prd,preprint,floatfix,superscriptaddress,nofootinbib,longbibliography]{revtex4-2}

\usepackage[T1]{fontenc}
\usepackage{lmodern}
\usepackage{amsmath,amssymb,amsfonts,mathtools,bm}
\usepackage{graphicx}
\usepackage{xcolor}
\usepackage{booktabs}
\usepackage{hyperref}
\usepackage{microtype}
\usepackage{physics}
\usepackage{enumitem}
\usepackage{array}
\usepackage{float}
\allowdisplaybreaks

\hypersetup{colorlinks=true,citecolor=blue,linkcolor=blue,urlcolor=blue}

\newcommand{\mpl}{M_{\rm Pl}}
\newcommand{\PR}{\mathcal P_{\mathcal R}}
\newcommand{\PT}{\mathcal P_T}

\begin{document}

\title{Inflation from a Weyl-flat null origin}
 \author{Malaika Arshad }
\email{malaikaarshad747@gmail.com }
 \affiliation{Department of Mathematics, Quaid-i-Azam University, Islamabad 45320, Pakistan}
\author{Jehanzad Zafar }
\email{jehanzadzafar@stu.xjtu.edu.cn}
 \affiliation{State Key Laboratory for Mechanical Behavior of Materials, School of Materials Science and Engineering, Xi’an Jiaotong University, Xi’an 710049, P. R. China}
  \affiliation{Department of Physics, Quaid-i-Azam University, Islamabad 45320, Pakistan}
\author{Ahdab Althukair}
\affiliation{ Department of Physics, College of Sciences, Princess Nourah bint Abdulrahman University, Riyadh 11671, Saudi Arabia}
\author{Bilal Ahmad}
\email{bilalahmad@emails.bjut.edu.cn}
\affiliation{Institute of Theoretical Physics, School of Physics and Optoelectronic Engineering, Beijing University of Technology, Beijing 100124, China}
\begin{abstract}
We show that a Weyl-flat null origin of inflation need not be in tension with present observations. For canonical single-field inflation, any background with $\epsilon(N)\to \epsilon_\infty\in(0,1)$ as $N\to\infty$ is asymptotically power-law, inherits the same Weyl-flat null past boundary, and reconstructs an exponential tail in field space. This identifies the origin as an asymptotic universality class rather than a rigid exact solution. We study a minimal deformation, $\epsilon(N)=\epsilon_\infty+(1-\epsilon_\infty)\left(\frac{N_0}{N+N_0}\right)^p$ with $p>1$, which preserves the asymptotic geometry, yields a smooth exit, and produces realistic finite-$N$ phenomenology. Solving the scalar and tensor mode equations directly in e-fold time, we find a viable corridor with $n_s$ in the Planck-preferred range and $r\sim10^{-3}-10^{-2}$, including reheating-compatible benchmarks. The result is a calculable single-field framework in which a Penrose-compatible Weyl-flat inflationary origin survives as a realistic and testable possibility.
\end{abstract}

\maketitle

\section{Introduction}

Inflation remains the most economical framework for explaining the near-flatness and near-homogeneity of the observed Universe while simultaneously generating an almost scale-invariant spectrum of primordial fluctuations~\cite{Guth:1980zm,Linde:1981mu,Albrecht:1982wi,Starobinsky:1980te,MukhanovChibisov:1981xt,Starobinsky:1982ee,Bardeen:1983qw}. Precision CMB data now impose a sharper standard: the scalar tilt is known to be red, the amplitude of primordial fluctuations is accurately measured, and the tensor sector is constrained strongly enough that broad classes of models can be compared directly with observation~\cite{Planck:2018inflation,BICEPKeck:2021gln,BICEPKeck:2024review,Calabrese:2025act,Wang:2024pgw}. Inflationary model building must therefore satisfy both conceptual and phenomenological criteria.

One longstanding conceptual concern is Penrose's Weyl-curvature hypothesis (WCH), according to which the initial state of the Universe should possess extremely low gravitational entropy, represented semiclassically by vanishing or strongly suppressed Weyl curvature near the initial boundary~\cite{Penrose:1979,Hu:2021}. This requirement is often taken to stand in tension with inflation. At the level of classical FRW backgrounds, however, that tension is not automatic. More broadly, attempts to formulate inflationary beginnings in terms of nontrivial boundary data already suggest that the structure of the past boundary and the dynamics of the observable inflationary phase need not be identified point by point~\cite{Hawking:1998bn}.

A particularly clear counterexample is already available in the literature~\cite{Damico:2022}. Exact power-law inflation generated by an exponential potential begins at a past null singularity while the metric remains exactly Friedmann-Robertson-Walker (FRW), and therefore conformally flat with identically vanishing Weyl tensor. At the classical background level, inflation may thus satisfy, rather than violate, a Penrose-compatible Weyl-flat origin. The familiar difficulty is instead phenomenological: the simplest power-law attractor predictions for $(n_s,r)$ are generally regarded as observationally strained~\cite{Lucchin:1984yf,Liddle:1994dx,Planck:2018inflation,BICEPKeck:2021gln}. A recent exact reanalysis of exponential-potential inflation further indicates that the full solution space is broader than the strict attractor treatment often emphasized in the literature~\cite{Yu:2026}.

The problem may therefore be stated more precisely. The issue is not whether a Weyl-compatible inflationary origin exists; exact power-law inflation already supplies such a background. The issue is whether that asymptotic origin can be preserved while the finite-$N$ dynamics are deformed in a controlled way so that the observable window yields an acceptable scalar tilt, a sufficiently small tensor amplitude, a graceful exit, and a plausible post-inflationary history. Equivalently, can one separate the \emph{geometric universality class} of the far past from the \emph{phenomenological requirements} of the CMB window within a minimal canonical single-field model?

That question is sharper than the familiar yes-or-no dispute over whether inflation ``solves'' the initial-conditions problem. The issue is the division between what the far past fixes and what the observable era is free to vary. In the present setting, the causal and conformal structure of the remote past is fixed, whereas the CMB window is shaped by a controlled finite-$N$ departure. The problem is therefore one of asymptotic inheritance: which properties of the origin remain imprinted in the observable universe, and which are diluted by the subsequent evolution.

The analysis below shows that this separation can be realized explicitly. The decisive point is that exact power-law inflation is not itself the invariant content. What survives is the far-past condition
\begin{equation}
    \epsilon(N)\to\epsilon_\infty\in(0,1)
    \qquad (N\to\infty),
\end{equation}
where $N$ denotes the number of e-folds remaining until the end of inflation. Once this criterion is isolated, the Weyl-flat null origin becomes an asymptotic property rather than an everywhere-exact constraint. This opens a broad class of deformations that preserve the asymptotic background structure while modifying the finite-$N$ dynamics that govern observable predictions.

We adopt the minimal construction capable of addressing this question. A single deformation of the Hubble-flow function $\epsilon(N)$ suffices to reconstruct the background, derive the scalar and tensor dynamics directly in e-fold time, and compute the spectra by numerical mode evolution rather than by relying on lowest-order slow-roll estimates. We then chart the viable corridor and incorporate reheating-aware pivot matching, which separates points that merely fall within the CMB ellipse from those that also admit a conventional thermal history~\cite{Liddle:2003as,Dai:2014jja,Martin:2014nya,Cook:2015vqa}. In this respect the framework is naturally aligned with recent studies that connect inflationary dynamics and post-inflationary sectors while retaining predictive control over late-time observables~\cite{Pirzada:2026dilaton,Pirzada:2026qcdaxion}.

The resulting class of \emph{asymptotically Weyl-flat inflationary models} consists of canonical Einstein-scalar backgrounds that are power-law and Weyl-flat in the far past, phenomenologically acceptable in the observable window, and smoothly terminating at finite $N$. The scope is correspondingly specific. We do not derive the quantum state from first principles, nor do we specify a UV-complete reheating sector. What is established is that the geometric mechanism identified at the background level survives within a realistic single-field model, and that the remaining uncertainties can be formulated as sharply defined problems. The result recasts a broad conceptual objection as a quantitative statement about how asymptotic geometry constrains inflationary observables.

\section{Geometric criterion for the null origin}

\subsection{Background equations and flow variable}

We consider a canonical scalar field minimally coupled to Einstein gravity in a spatially flat FRW spacetime,
\begin{equation}
    ds^2=-dt^2+a^2(t)d\vb{x}^2,
\end{equation}
with background equations
\begin{align}
    3\mpl^2H^2 &= \frac12\dot\phi^2+V(\phi),
    \label{eq:Friedmann1}\\
    -2\mpl^2\dot H &= \dot\phi^2,
    \label{eq:Friedmann2}\\
    \ddot\phi+3H\dot\phi+V_{,\phi} &=0.
    \label{eq:KGeq}
\end{align}
The dynamics may be encoded in the Hamilton-Jacobi form~\cite{Salopek:1990jq,Liddle:1994dx},
\begin{equation}
    \dot\phi=-2\mpl^2H_{,\phi},
    \qquad
    V(\phi)=3\mpl^2H^2-2\mpl^4H_{,\phi}^2,
    \label{eq:HJ}
\end{equation}
which makes clear that once the Hubble function is known as a function of field space, the entire background follows.

For the present problem it is more useful to work with the number of e-folds \emph{remaining} until the end of inflation,
\begin{equation}
    N\equiv\ln\frac{a_{\rm end}}{a},
\end{equation}
so that $N=0$ at the end of inflation and $N\to\infty$ in the far past. Since $\dot N=-H$, one obtains the exact flow equations
\begin{equation}
    \frac{d\ln H}{dN}=\epsilon,
    \qquad
    \frac{d\phi}{dN}=-\mpl\sqrt{2\epsilon},
    \label{eq:reconeqs}
\end{equation}
where
\begin{equation}
    \epsilon\equiv -\frac{\dot H}{H^2}
    =\frac{1}{2\mpl^2}\left(\frac{d\phi}{dN}\right)^2.
\end{equation}
Equation~\eqref{eq:reconeqs} summarizes the reconstruction. The first relation states that the logarithmic growth of the Hubble scale toward the past is controlled pointwise by $\epsilon(N)$. The second shows that the field excursion is the integral of $\sqrt{2\epsilon}$. In this formulation, $\epsilon(N)$ simultaneously fixes the background clock, the reconstruction variable, and the asymptotic causal structure of the spacetime. This is also the sense in which the present setup belongs to the broader reconstruction program, but with the asymptotic causal question built directly into the flow variable from the outset~\cite{Lin:2015fqa}.

For later use we define the Hubble-flow hierarchy~\cite{Kinney:2002qn,Schwarz:2001vv},
\begin{equation}
    \epsilon_1\equiv\epsilon,
    \qquad
    \epsilon_2\equiv\frac{d\ln\epsilon_1}{dN},
    \qquad
    \epsilon_3\equiv\frac{d\ln\epsilon_2}{dN}.
    \label{eq:hubblehierarchy}
\end{equation}
The reconstructed potential along the background is
\begin{equation}
    V(N)=\mpl^2H^2(N)[3-\epsilon(N)],
    \label{eq:VN}
\end{equation}
and its exact logarithmic slope can be written as
\begin{equation}
    \mpl\frac{V_{,\phi}}{V}=-\sqrt{\frac{\epsilon_1}{2}}\left[2-\frac{\epsilon_2}{3-\epsilon_1}\right].
    \label{eq:Vslopeexact}
\end{equation}
Equation~\eqref{eq:Vslopeexact} makes the logic transparent: the steepness of the reconstructed potential is governed not only by the magnitude of $\epsilon_1$ but also by the local drift encoded in $\epsilon_2$. That extra leverage is what preserves an exponential tail in the far past while allowing a much gentler slope near horizon exit.

\subsection{Asymptotic proposition}

The relevant geometric statement can be formulated as follows.

\paragraph*{Proposition.}
Assume that a canonical single-field flat-FRW inflationary background satisfies
\begin{equation}
    \epsilon(N)\xrightarrow[N\to\infty]{}\epsilon_\infty,
    \qquad 0<\epsilon_\infty<1,
    \label{eq:epslimit}
\end{equation}
and that $\epsilon(N)-\epsilon_\infty$ is integrable for sufficiently large $N$. Then:
\begin{enumerate}[leftmargin=1.8em]
    \item the far past is asymptotic to power-law inflation with constant $\epsilon_\infty$;
    \item the reconstructed potential acquires an exponential tail,
    $V(\phi)=V_0e^{-\sqrt{2\epsilon_\infty}\,\phi/\mpl}[1+o(1)]$ as $\phi\to-\infty$;
    \item the background Weyl tensor vanishes identically, and the causal character of the past boundary is the same null boundary found in exact power-law inflation~\cite{Damico:2022}.
\end{enumerate}

\paragraph*{Proof.}
Integrating Eq.~\eqref{eq:reconeqs} gives
\begin{align}
    \ln H(N)&=\ln H_\infty+\epsilon_\infty N+o(1),
    \label{eq:lnHasym}\\
    \phi(N)&=\phi_\infty-\mpl\sqrt{2\epsilon_\infty}\,N+o(N),
    \label{eq:phiasym}
\end{align}
where the integrability assumption ensures that the correction to $\ln H$ stays bounded at large $N$. Eliminating $N$ between these expressions yields
\begin{equation}
    H(\phi)=H_0\exp\!\bigg[-\sqrt{\frac{\epsilon_\infty}{2}}\,\frac{\phi}{\mpl}\bigg][1+o(1)],
\end{equation}
and therefore, by Eq.~\eqref{eq:VN},
\begin{equation}
    V(\phi)=V_0\exp\!\bigg[-\sqrt{2\epsilon_\infty}\,\frac{\phi}{\mpl}\bigg][1+o(1)],
    \qquad \phi\to-\infty.
    \label{eq:VasymptoticGeneral}
\end{equation}
To identify the metric asymptotics, note that $a=e^{-N}$ and $dt=-dN/H$. Equation~\eqref{eq:lnHasym} implies $dt\propto e^{-\epsilon_\infty N}dN$, hence $t\propto e^{-\epsilon_\infty N}$ as $N\to\infty$, so that
\begin{equation}
    a(t)\propto t^{1/\epsilon_\infty},
    \qquad
    H(t)\sim \frac{1}{\epsilon_\infty t},
    \qquad t\to0^+.
    \label{eq:powerlawpast}
\end{equation}
This is exactly the power-law form associated with an exponential potential~\cite{Lucchin:1984yf,Liddle:1994dx,Ratra:1984ia}. Because the metric remains FRW for all $t$, the Weyl tensor vanishes identically. Finally, the conformal time behaves as
\begin{equation}
    \eta\sim -\int^\infty dN\, e^{-(1-\epsilon_\infty)N}\to -\infty,
    \qquad 0<\epsilon_\infty<1,
\end{equation}
so the past boundary has the same null character as in the exact power-law solution. \hfill$\square$

The proposition pinpoints the real constraint. Exact power-law evolution over the whole history is unnecessary. What matters is the asymptotic approach to a constant $\epsilon_\infty<1$. Once that holds, the far past is fixed by the exponential tail, whereas the CMB observables are controlled by the finite-$N$ departure from it.

This distinction deserves emphasis. The far past does not fix the inflationary phenomenology point by point; it defines a universality class. The null boundary, the Weyl-flat FRW geometry, and the exponential asymptotic tail are the structures that survive as $N\to\infty$. By contrast, the scalar tilt, tensor amplitude, and reheating map are set by how the background departs from that asymptotic regime during the last few dozen e-folds. In this sense the model separates origin data from observational data. That separation is what makes Penrose-type questions meaningful without forcing the observable universe to replicate exact power-law inflation.

The proposition is equally consistent with past geodesic incompleteness. A background can be Weyl-flat and still be past-incomplete, in line with the standard incompleteness theorem for inflationary spacetimes~\cite{Borde:2001nh}. The present construction does not evade that theorem. It makes a narrower point: low Weyl curvature, null-origin structure, and geodesic completeness are distinct questions, and treating them as if they were the same only blurs what inflation actually establishes.

\section{Reconstruction of the flow family}

\subsection{A minimal flow ansatz}

The minimal deformation we consider is
\begin{equation}
    \epsilon(N)=\epsilon_\infty+(1-\epsilon_\infty)\left(\frac{N_0}{N+N_0}\right)^p,
    \qquad p>1.
    \label{eq:epsansatz}
\end{equation}
This ansatz is as spare as one can reasonably make it. The parameter $\epsilon_\infty$ fixes the asymptotic power-law index and therefore the slope of the exponential tail. The scale $N_0$ sets the e-fold interval over which the background begins to peel away from the asymptotic branch, while $p$ controls how abruptly that peeling occurs. By construction,
\begin{equation}
    \epsilon(0)=1,
\end{equation}
so the model exits inflation smoothly at $N=0$ without an auxiliary waterfall field or a separate postulated termination mechanism.

The analytic Hubble reconstruction follows immediately from Eq.~\eqref{eq:reconeqs},
\begin{equation}
    \ln\frac{H(N)}{H_{\rm end}}
    =\epsilon_\infty N+
    \frac{(1-\epsilon_\infty)N_0^p}{1-p}
    \left[(N+N_0)^{1-p}-N_0^{1-p}\right].
    \label{eq:Hanalytic}
\end{equation}
The field profile and potential are then obtained exactly by quadrature,
\begin{align}
    \phi(N)-\phi_{\rm end} &= -\mpl\int_0^N d\tilde N\,\sqrt{2\epsilon(\tilde N)},
    \label{eq:phiquadrature}\\
    V(N) &= \mpl^2H^2(N)[3-\epsilon(N)].
    \label{eq:Vquadrature}
\end{align}
Because $d\phi/dN<0$, the field rolls monotonically from large negative values in the asymptotic past to a finite endpoint at inflation's end. The trajectory is therefore single-valued throughout, with no turning point or branch ambiguity inserted by hand.

For the profile in Eq.~\eqref{eq:epsansatz}, the next Hubble-flow functions are
\begin{align}
    \epsilon_2(N) &= -\frac{p(1-\epsilon_\infty)N_0^p}{(N+N_0)^{p+1}\epsilon(N)},
    \label{eq:eps2model}\\
    \frac{d\epsilon_2}{dN} &= \frac{p(1-\epsilon_\infty)N_0^p}{(N+N_0)^{p+2}\epsilon^2}
    \left[(p+1)\epsilon(N)-\frac{p(1-\epsilon_\infty)N_0^p}{(N+N_0)^p}\right].
    \label{eq:deps2model}
\end{align}
Since $\epsilon_2<0$, the red tilt in the observable window need not come from a large $\epsilon_1$. It is generated by the controlled drift away from the asymptotic branch. That drift is the essential phenomenological handle of the model.

\begin{figure}[t]
    \includegraphics[width=0.82\textwidth]{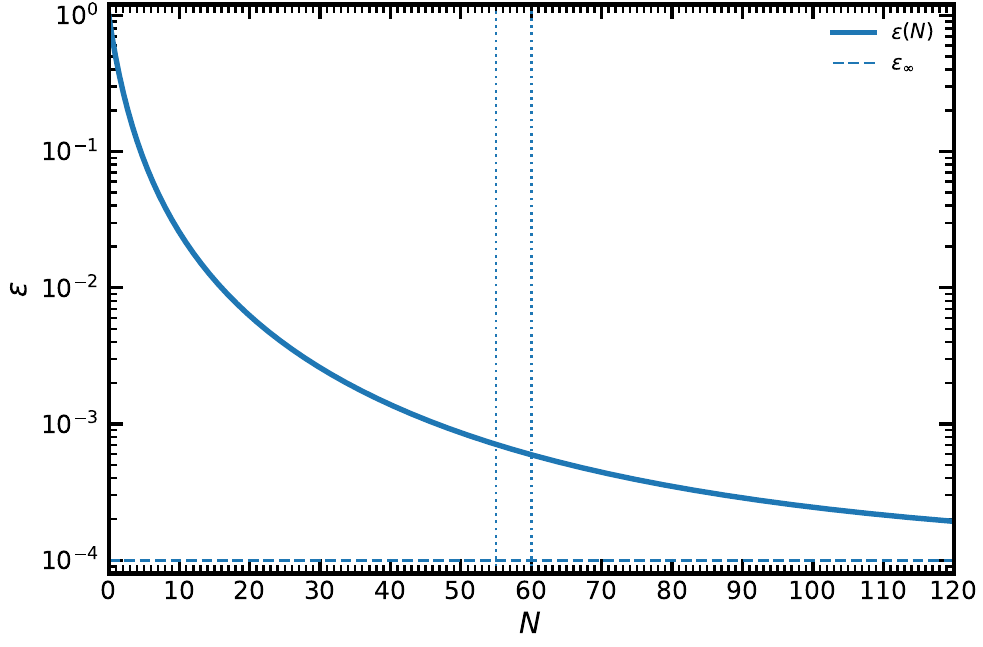}
    \caption{Flow profile for benchmark A, $(\epsilon_\infty,N_0,p)=(10^{-4},3,5/2)$. The asymptotic value $\epsilon_\infty$ fixes the power-law, Weyl-flat past, while the smooth rise to $\epsilon(0)=1$ closes inflation without an additional exit sector.}
    \label{fig:eps}
\end{figure}

\begin{figure}[t]
    \includegraphics[width=0.82\textwidth]{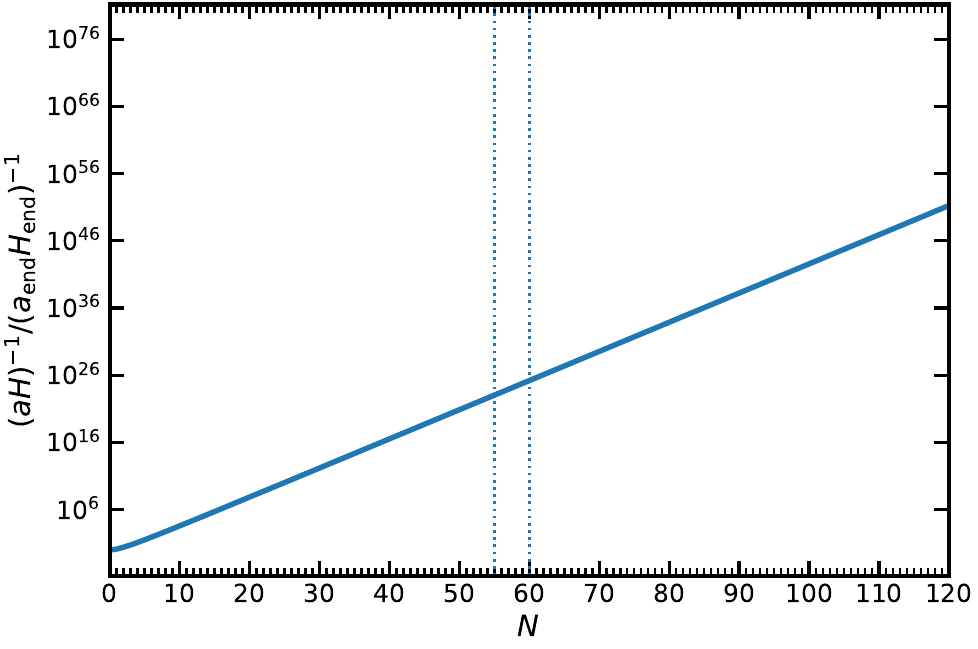}
    \caption{Comoving Hubble radius for benchmark A. Its monotonic decrease confirms that accelerated expansion persists throughout the reconstructed background until the smooth exit at $N=0$.}
    \label{fig:ah}
\end{figure}

A useful diagnostic is the equation-of-state parameter,
\begin{equation}
    w_\phi\equiv\frac{p_\phi}{\rho_\phi}=-1+\frac{2\epsilon}{3},
    \label{eq:wphi}
\end{equation}
which interpolates from $w_\infty=-1+2\epsilon_\infty/3$ in the far past to $w_\phi=-1/3$ at the end of inflation. The ansatz therefore has a clear physical interpretation: it describes a background that begins arbitrarily close to a deformed de Sitter state with fixed but nonzero $\epsilon_\infty$, lingers on a broad slow-roll shelf, and then steepens sufficiently to terminate inflation.

\subsection{Asymptotic tail and field excursion}

The asymptotic structure is immediate from Eq.~\eqref{eq:Hanalytic}. For $p>1$,
\begin{equation}
    \ln H(N)=\epsilon_\infty N+C_H+\mathcal O\!\big(N^{1-p}\big),
    \qquad N\to\infty,
\end{equation}
with finite constant $C_H$. Likewise,
\begin{equation}
    \phi(N)=\phi_\infty-\mpl\sqrt{2\epsilon_\infty}\,N+\mathcal O\!\big(N^{1-p}\big),
\end{equation}
so the potential approaches a pure exponential with power-suppressed corrections. The model is therefore not merely inspired by power-law inflation; it is asymptotically identical to that universality class in the far past.

\begin{figure}[t]
    \includegraphics[width=0.82\textwidth]{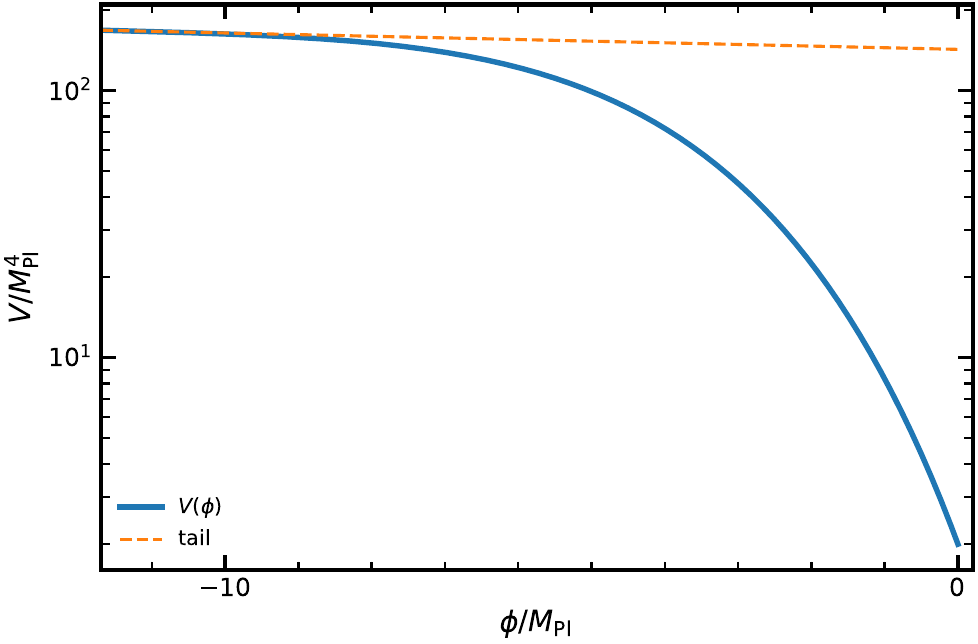}
    \caption{Reconstructed potential for benchmark A. The dashed line is the asymptotic exponential fit dictated by the proposition. The observable shelf and the late-time steepening are generated dynamically by the flow ansatz rather than imposed directly in field space.}
    \label{fig:potential}
\end{figure}

At finite $N$, however, the same reconstruction generates a qualitatively different field-space potential. Figure~\ref{fig:potential} shows the result for benchmark A. The left tail is exponentially straight, as required by the proposition. Around the observable regime, the potential develops a noticeably shallower shelf, and near the end of inflation it steepens again. These three regimes arise from a single smooth flow profile in $N$-space rather than from a piecewise potential imposed directly in field space.

This field-space representation makes the separation of roles explicit. The far-left tail is fixed almost entirely by $\epsilon_\infty$, whereas the CMB observables depend on the local values of $\epsilon_1$ and $\epsilon_2$ near horizon exit. The origin and the observable window are therefore linked, but not rigidly locked together. That limited correlation is exactly what allows the Weyl-flat power-law past to survive without importing the full phenomenological burden of exact power-law inflation.

Viewed in field space, the reconstructed potential naturally falls into three regimes: an exponential tail that fixes the causal character of the past boundary, an intermediate shelf that governs the observable perturbations, and a late steepening that ends inflation. This three-part structure carries information beyond a mere fit to $(n_s,r)$. It indicates that phenomenological freedom can be confined to a finite interval in e-fold time while the asymptotic sector remains fixed. The same perspective is relevant more broadly whenever ultraviolet reasoning constrains the tail of the potential rather than the full observable trajectory~\cite{Martin:2013tda,Khan:2023waterfall}.

The field excursion itself is moderate in the sense relevant for large-field slow-roll models, but not parametrically tiny. Direct integration of Eq.~\eqref{eq:phiquadrature} gives
\begin{equation}
    \Delta\phi_A\simeq 9.13\,\mpl,
    \qquad
    \Delta\phi_B\simeq 7.53\,\mpl,
\end{equation}
for the two benchmarks introduced below. This is consistent with a tensor amplitude in the $r\sim 10^{-2}$ range rather than with an exponentially suppressed signal. The model is therefore best viewed as a controlled large-field construction with a protected asymptotic tail, rather than as an ultra-small-field scenario.

\section{Exact perturbations and observables}

\subsection{Mode equations in e-fold time}

A broad survey of the model space can be performed with the first-order Hubble-flow expressions
\begin{equation}
    n_s-1\simeq -2\epsilon_1+\epsilon_2,
    \qquad
    r\simeq 16\epsilon_1,
    \label{eq:firstorderobs}
\end{equation}
but a precision benchmark should not rest solely on that approximation. We therefore solve the scalar and tensor mode equations directly~\cite{Mukhanov:1985rz,Sasaki:1986hm,BunchDavies:1978yq,Stewart:1993bc,Gong:2001he,Ringeval:2007am}.

For the canonically normalized scalar mode variable $u_k$, the conformal-time equation is
\begin{equation}
    u_k''+\left(k^2-\frac{z''}{z}\right)u_k=0,
    \label{eq:MSeta}
\end{equation}
where the prime denotes $d/d\eta$. In e-fold time one has $d/d\eta=-aH\,d/dN$, so Eq.~\eqref{eq:MSeta} becomes
\begin{equation}
    u_{k,NN}-(1-\epsilon_1)u_{k,N}+\left[\frac{k^2}{(aH)^2}-\Omega\right]u_k=0,
    \label{eq:modeeqN}
\end{equation}
with
\begin{equation}
    \Omega\equiv y_N+y^2-(1-\epsilon_1)y,
    \qquad y\equiv\frac{d\ln z}{dN}.
    \label{eq:OmegaDef}
\end{equation}
Equation~\eqref{eq:modeeqN} puts the exact background and the perturbations on the same clock. Because the reconstruction itself is formulated in $N$, the mode evolution can be followed without switching variables midstream. That is more than a technical convenience: the same parameter that measures distance from the end of inflation also measures distance from the asymptotic origin, so the link between early-time geometry and late-time observables remains explicit throughout.

For scalar modes, $z_s=a\sqrt{2\epsilon_1}\,\mpl$, so
\begin{equation}
    y_s=-1+\frac12\epsilon_2,
    \label{eq:ys}
\end{equation}
and
\begin{equation}
    \Omega_s=2-\epsilon_1-\frac32\epsilon_2+\frac12\epsilon_1\epsilon_2+\frac14\epsilon_2^2+\frac12\epsilon_2\epsilon_3.
    \label{eq:Omegas}
\end{equation}
For tensor modes, $z_t=a\mpl$, which gives the especially simple result
\begin{equation}
    \Omega_t=2-\epsilon_1.
    \label{eq:Omegat}
\end{equation}
A direct symbolic differentiation confirms Eqs.~\eqref{eq:Omegas} and \eqref{eq:Omegat}.

We impose the standard adiabatic vacuum initial condition deep inside the horizon,
\begin{equation}
    u_k(\eta_i)=\frac{e^{-ik\eta_i}}{\sqrt{2k}},
    \qquad
    \frac{k}{aH}\bigg|_{\eta_i}=100,
    \label{eq:BD}
\end{equation}
and evolve each mode until freeze-out near $N\simeq0.05$. The power spectra are then
\begin{equation}
    \PR(k)=\frac{k^3}{2\pi^2}\left|\frac{u_k}{z_s}\right|^2,
    \qquad
    \PT(k)=\frac{4k^3}{\pi^2}\left|\frac{u_k}{z_t}\right|^2.
    \label{eq:powerdefs}
\end{equation}
The spectral observables are extracted from local fits around the pivot,
\begin{equation}
    n_s-1=\frac{d\ln\PR}{d\ln k},
    \qquad
    \alpha_s=\frac{dn_s}{d\ln k},
    \qquad
    n_t=\frac{d\ln\PT}{d\ln k},
    \qquad
    r=\frac{\PT}{\PR}\bigg|_{k_*}.
    \label{eq:obsdefs}
\end{equation}
With this setup, the benchmark values are direct predictions of the reconstructed background rather than residues of a lowest-order approximation or a particular pivot convention.

\subsection{Benchmark A: spectra and consistency}

We first present benchmark A,
\begin{equation}
    \text{benchmark A:}
    \qquad
    (\epsilon_\infty,N_0,p,N_*)=(10^{-4},3,5/2,60).
\end{equation}
which is best interpreted as a benchmark selected to lie well within the CMB-allowed region. Numerical integration of the mode equations yields
\begin{align}
    n_s &= 0.96633,
    & r &= 9.29\times10^{-3},
    \nonumber\\
    \alpha_s &= -7.69\times10^{-4},
    & n_t &= -1.20\times10^{-3},
    \label{eq:benchAobs}
\end{align}
with inflationary scale
\begin{equation}
    V_*^{1/4}=1.00\times10^{16}\,{\rm GeV}.
    \label{eq:benchAenergy}
\end{equation}
The leading Hubble-flow estimate gives $n_s\simeq0.96580$ and $r\simeq9.52\times10^{-3}$, so the approximation already tracks the exact result well. This agreement shows that the benchmark lies in a regime where slow-roll intuition remains reliable, even though the final numbers are extracted from the exact mode evolution.

\begin{figure}[t]
    \includegraphics[width=0.82\textwidth]{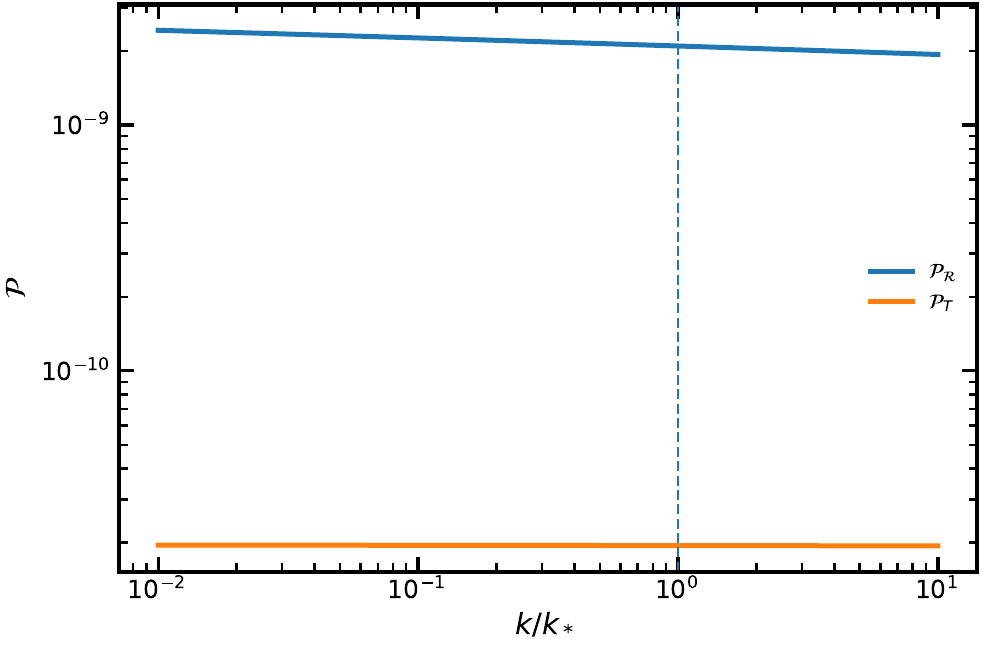}
    \caption{Exact scalar and tensor spectra for benchmark A obtained by integrating the mode equations in e-fold time. The spectra are close to local power laws across the CMB window, but not exactly so; a direct numerical extraction of $n_s$ and $\alpha_s$ is therefore cleaner than a purely analytic fit.}
    \label{fig:spectra}
\end{figure}

\begin{table}[t]
    \caption{Representative benchmarks. Benchmark A is a point chosen to illustrate viability in the CMB plane. Benchmark B is a nearby point for which the same asymptotic mechanism also fits a conventional reheating history. The field excursion is obtained by direct integration of Eq.~\eqref{eq:phiquadrature}.}
    \label{tab:benchmarks}
    \begin{ruledtabular}
    \begin{tabular}{lcc}
        Quantity & Benchmark A & Benchmark B \\
        \midrule
        $(\epsilon_\infty,N_0,p,N_*)$ & $(10^{-4},3,5/2,60)$ & $(10^{-4},2.4,5/2,55)$ \\
        $n_s$ (exact) & $0.96633$ & $0.96572$ \\
        $r$ (exact) & $9.29\times10^{-3}$ & $7.14\times10^{-3}$ \\
        $\alpha_s$ (exact) & $-7.69\times10^{-4}$ & $-9.27\times10^{-4}$ \\
        $n_t$ (exact) & $-1.20\times10^{-3}$ & $-9.24\times10^{-4}$ \\
        $n_s$ (first order) & $0.96580$ & $0.96505$ \\
        $r$ (first order) & $9.52\times10^{-3}$ & $7.32\times10^{-3}$ \\
        $V_*^{1/4}$ & $1.00\times10^{16}\,$GeV & $9.40\times10^{15}\,$GeV \\
        $V_{\rm end}^{1/4}$ & $3.72\times10^{15}\,$GeV & $4.24\times10^{15}\,$GeV \\
        $\Delta\phi/\mpl$ & $9.13$ & $7.53$ \\
    \end{tabular}
    \end{ruledtabular}
\end{table}

Several consistency checks are immediate. The tensor tilt remains close to the canonical single-field consistency estimate $n_t\simeq-r/8$, but is not imposed by hand because it is extracted from the exact tensor spectrum. The running is small and negative, as expected in a smooth single-field model without sharp features. The inflationary scale lies near the conventional GUT scale while remaining within current tensor bounds.

Benchmark A already exhibits the central structural result: a Weyl-flat null origin does not compel the observable window to inherit the rigid predictions of exact power-law inflation. The asymptotic geometry survives intact, while the finite-$N$ deformation opens a CMB window consistent with current data.

At this stage the construction ceases to be purely formal. The observable sector is not recovered by abandoning the geometric premise, but by determining how a transient shelf can be attached to an asymptotically fixed tail. The relevant question is therefore not whether a Weyl-flat origin can be rendered compatible with data in the abstract, but how much phenomenological latitude can be generated before the finite-$N$ departure ceases to qualify as a mild deformation of the same asymptotic universe. The benchmarks provide concrete coordinates within that broader question.

\section{Phenomenology and reheating}
\label{sec:pheno}

\subsection{The viable corridor and benchmark B}
\label{sec:scan}

The benchmark is not isolated. To expose the width of the viable region we scan
\begin{equation}
    N_0\in[2,4.5],
    \qquad
    p\in[2.1,3.2],
    \qquad
    \epsilon_\infty=10^{-4},
    \label{eq:scanrange}
\end{equation}
using the first-order Hubble-flow expressions evaluated at $N_*=55$. This level of approximation is appropriate for a global map: exact mode integration is reserved for benchmarks, while a flow-based survey is sufficient to identify the overall viable corridor.

\begin{figure}[t]
    \includegraphics[width=0.82\textwidth]{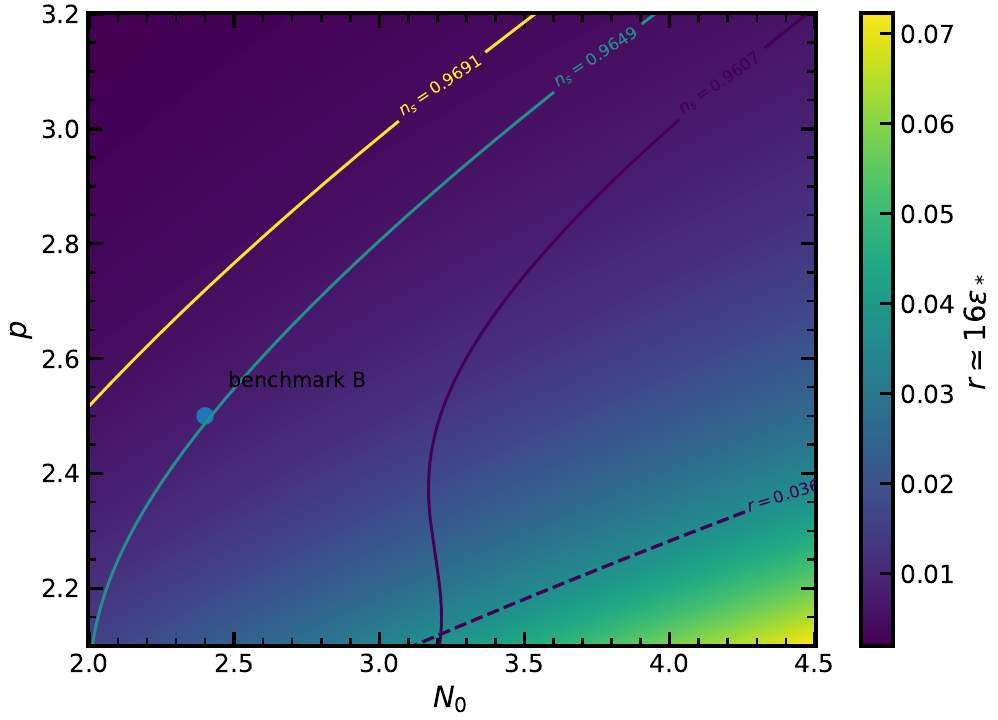}
    \caption{Parameter-space structure for $\epsilon_\infty=10^{-4}$ at $N_*=55$. The color scale shows $r\simeq16\epsilon_*$, while the contours track the Planck central value of $n_s$ and its $1\sigma$ band. Benchmark B sits comfortably inside the viable corridor.}
    \label{fig:scan}
\end{figure}

Figure~\ref{fig:scan} shows that a substantial band in the $(N_0,p)$ plane yields values of $n_s$ inside the Planck-preferred region while keeping $r$ below the BK18 upper limit~\cite{Planck:2018inflation,BICEPKeck:2021gln}. In the scanned domain, the $68\%$-compatible corridor spans approximately
\begin{equation}
    3.6\times10^{-3}\lesssim r\lesssim 3.6\times10^{-2},
\end{equation}
which is already phenomenologically significant. The model does not merely squeeze into the data at one tuned point; it occupies an extended and observationally nontrivial tensor corridor.

That corridor has a clear physical interpretation. The asymptotic Weyl-flat requirement is restrictive enough to prevent the model from degenerating into an arbitrarily flexible reconstruction, yet not so restrictive that it collapses to a single observational target. In practice, the framework predicts a band of tensor amplitudes that can be sharpened by future B-mode limits. A substantial downward shift in the allowed upper bound on $r$ would therefore do more than exclude a benchmark; it would begin to test the broader strategy of attaching the CMB window to a power-law asymptotic past through a smooth finite-$N$ deformation. This is consistent with current data analyses, where reheating and late-time likelihoods often compress apparently broad inflationary families into comparatively narrow observational corridors~\cite{Ellis:2025act,Haque:2025plateau}.

\begin{figure}[t]
    \includegraphics[width=0.82\textwidth]{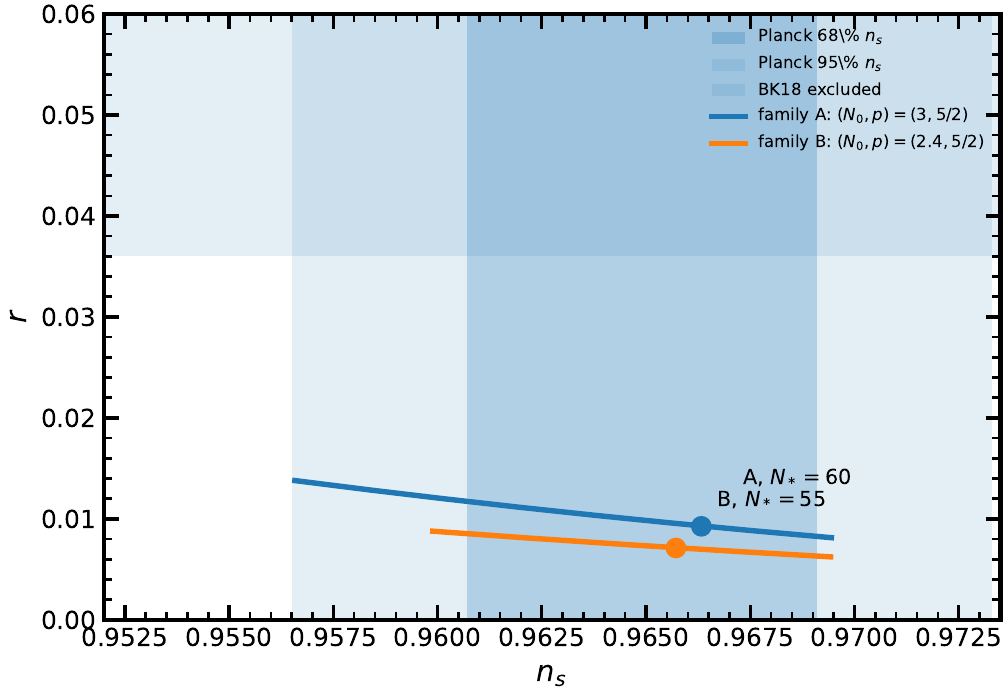}
    \caption{Observable families generated by varying $N_*$. Family A corresponds to $(N_0,p)=(3,5/2)$ and family B to $(2.4,5/2)$, both with $\epsilon_\infty=10^{-4}$. The exact benchmark points are superimposed.}
    \label{fig:constraints}
\end{figure}

The retuned benchmark that we will use as the phenomenological reference point is
\begin{equation}
    \text{benchmark B:}
    \qquad
    (\epsilon_\infty,N_0,p,N_*)=(10^{-4},2.4,5/2,55).
\end{equation}
Its exact observables are
\begin{equation}
    n_s=0.96572,
    \qquad
    r=7.14\times10^{-3},
    \qquad
    \alpha_s=-9.27\times10^{-4},
    \qquad
    n_t=-9.24\times10^{-4}.
\end{equation}
This point lies close to the Planck central value for $n_s$ and comfortably below current tensor bounds. Figure~\ref{fig:constraints} shows the families generated by varying $N_*$ around the two benchmark profiles. The allowed region is therefore a corridor rather than an isolated coincidence.

Qualitatively, increasing $p$ at fixed $N_0$ sharpens the transition away from the asymptotic branch and typically lowers $r$. Increasing $N_0$ at fixed $p$ shifts the deformation closer to the CMB window and tends to increase $\epsilon_*$, thereby raising $r$. Meanwhile, the observables remain only weakly sensitive to the tiny value of $\epsilon_\infty$ so long as horizon crossing occurs on the deformed shelf rather than deep in the asymptotic tail. The geometric origin is therefore controlled primarily by $\epsilon_\infty$, whereas the CMB window is controlled mainly by $(N_0,p)$.

\subsection{Reheating-aware pivot map}
\label{sec:reheating}

The number of e-folds between horizon exit and the end of inflation is fixed by the post-inflationary expansion history rather than chosen arbitrarily~\cite{Liddle:2003as,Dai:2014jja,Martin:2014nya,Cook:2015vqa}. In a framework like the present one, where $n_s$ and $r$ drift visibly as $N_*$ varies, this point is not a technical aside. It is part of the physical prediction.

We therefore parameterize reheating by an effective constant equation-of-state parameter $w_{\rm re}$ between the end of inflation and thermalization,
\begin{equation}
    \rho_{\rm re}=\rho_{\rm end}\exp\left[-3(1+w_{\rm re})N_{\rm re}\right],
    \label{eq:rhoreheat}
\end{equation}
with thermal density
\begin{equation}
    \rho_{\rm re}=\frac{\pi^2}{30}g_{\rm re}T_{\rm re}^4.
    \label{eq:rhorad}
\end{equation}
Using entropy conservation,
\begin{equation}
    g_{s,\rm re}T_{\rm re}^3a_{\rm re}^3=\frac{43}{11}T_0^3a_0^3,
\end{equation}
and the pivot identity $k_*=a_*H_*=a_{\rm end}e^{-N_*}H_*$, one obtains the standard reheating relations
\begin{equation}
    N_{\rm re}=\frac{4}{1-3w_{\rm re}}\Bigg[
    -N_* - \ln\frac{k_*}{a_0H_0} + \ln\frac{H_*}{H_0}
    + \ln T_0
    + \frac13\ln\frac{43}{11g_{s,\rm re}}
    - \frac14\ln\frac{30}{\pi^2 g_{\rm re}}
    - \frac14\ln\rho_{\rm end}
    \Bigg],
    \label{eq:Nreformula}
\end{equation}
\begin{equation}
    T_{\rm re}=\left(\frac{30\rho_{\rm end}}{\pi^2 g_{\rm re}}\right)^{1/4}
    \exp\left[-\frac34(1+w_{\rm re})N_{\rm re}\right].
    \label{eq:Treformula}
\end{equation}
These relations turn the inflationary solution into a post-inflationary diagnostic. They separate benchmark points that only land in the right region of the CMB plane from those that can also be embedded in a plausible thermal history.

Applying Eq.~\eqref{eq:Nreformula} to benchmark A identifies a nontrivial separation: for representative values $w_{\rm re}=-0.2$, $0$, and $0.25$, one obtains negative $N_{\rm re}$. Benchmark A therefore remains admissible as a \emph{CMB benchmark}, but not as a self-consistent \emph{reheating benchmark}. This does not indicate any inconsistency of the construction; it shows that the framework resolves the distinction between CMB viability and post-inflationary viability at the level of explicit diagnostics.

Benchmark B resolves this cleanly. For $(\epsilon_\infty,N_0,p,N_*)=(10^{-4},2.4,5/2,55)$, the reheating diagnostics become
\begin{align}
    w_{\rm re}=-0.2 &: & N_{\rm re}=3.34, &\qquad T_{\rm re}=2.34\times10^{14}\,{\rm GeV},
    \nonumber\\
    w_{\rm re}=0 &: & N_{\rm re}=5.35, &\qquad T_{\rm re}=3.15\times10^{13}\,{\rm GeV},
    \nonumber\\
    w_{\rm re}=0.25 &: & N_{\rm re}=21.4, &\qquad T_{\rm re}=3.35\times10^{6}\,{\rm GeV}.
    \label{eq:benchBreheat}
\end{align}
These numbers are entirely conventional by the standards of single-field reheating analyses. Their significance is that the same asymptotically Weyl-flat mechanism can be embedded in a standard post-inflationary history~\cite{Ellis:2025act}.

\begin{figure}[t]
    \includegraphics[width=0.82\textwidth]{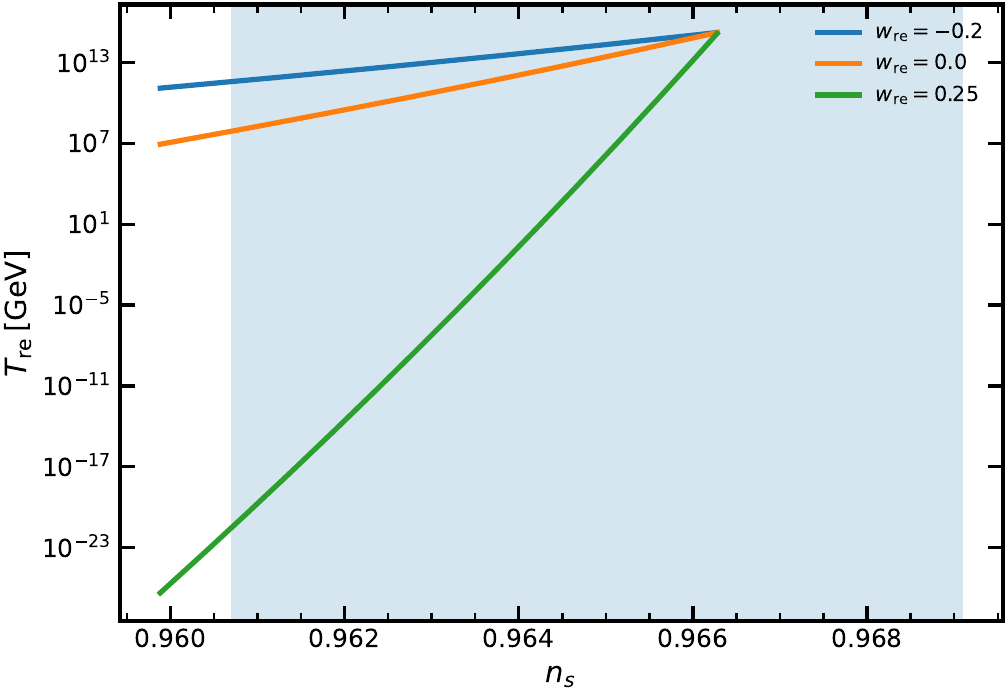}
    \caption{Reheating map for benchmark family B. Each curve is obtained by varying $N_*$ and converting the result into a reheating temperature via Eqs.~\eqref{eq:Nreformula} and \eqref{eq:Treformula}. The shaded vertical band denotes the Planck $68\%$ interval for $n_s$.}
    \label{fig:reheating}
\end{figure}

Figure~\ref{fig:reheating} displays the resulting map. Across a wide range of $w_{\rm re}$, the Planck-favored region of $n_s$ corresponds to positive reheating durations and plausible reheating temperatures. The model is therefore not only geometrically and observationally viable; it also passes the first nontrivial post-inflationary consistency check, in line with recent reheating-aware comparisons of inflationary model classes~\cite{Ellis:2025act}.

This reheating map also sharpens an interpretive point that is easy to miss when one quotes only $(n_s,r)$. Moving along the family does not merely shift a point in abstract parameter space; it changes the cosmic history between horizon exit and thermalization. In that sense the reheating analysis links the asymptotic origin to a fuller cosmological narrative. That is what makes the reconstruction physical rather than merely kinematic.

\subsection{Broader phenomenological relevance}

The interest of the framework does not stop at the existence of a viable region. More significantly, it reshapes how one organizes the link between early-universe geometry and inflationary observables.

At a conceptual level, the construction makes the initial-state question operational. Discussions of Penrose-compatible beginnings often remain schematic: low Weyl curvature is invoked as a desideratum, while inflation is cast as either ally or adversary. Here the issue becomes calculable. A Weyl-flat null origin is imposed as an asymptotic condition on the flow, and the measurable consequences are followed through the finite-$N$ deformation that connects that origin to the CMB window. One can then ask, quantitatively, how much of the asymptotic structure survives and where the data cut most deeply.

From the standpoint of classification, the model suggests a different notion of inflationary universality. Many successful constructions are classified by the local shape of the potential near horizon exit~\cite{Martin:2013tda,Ellis:2025act,Haque:2025plateau}. Here that logic is partly inverted: the far-past asymptotics define the universality class, while the observable era is governed by a controlled departure from it. The taxonomy therefore depends not only on the slope and curvature near the pivot, but also on the causal and conformal data encoded in the asymptotic flow. From this perspective, the present model is notable because it preserves an asymptotically fixed origin while still generating a nontrivial observable shelf.

At the observational level, the tensor sector is neither driven back to the exact power-law level nor diluted into invisibility, but instead occupies a corridor at $r\sim10^{-3}-10^{-2}$. That range is testable in a substantive way. Stronger B-mode bounds would progressively carve away the allowed shelf structure, while a future detection in this interval would provide a natural observational target~\cite{BICEPKeck:2024review,Wang:2024pgw,Hertig:2024so}. The phenomenology is therefore not an afterthought attached to a geometric idea; it is the arena in which that geometric premise becomes exposed to data.

At the level of model extension, the analysis points naturally toward scenarios in which the post-inflationary sector is dynamically active rather than a passive sink, and in which localized structure in field space generates additional phenomenology without compromising control of the inflationary trajectory~\cite{Pirzada:2026axionsu2,Pirzada:2026patisalam,Khan:2024gutgw,Khan:2023waterfall}. One may then ask whether anisotropic or weakly inhomogeneous deformations inherit the same asymptotic discipline, whether non-adiabatic initial states can be organized without spoiling the null-origin picture, and whether microscopic sectors can generate the required finite-$N$ shelves without destabilizing the exponential tail. These are not generic placeholders. They are the concrete points at which the present mechanism could either acquire a more compelling microphysical interpretation or encounter its sharpest obstructions. In that sense, the construction is important not only because it closes one loophole in a conceptual objection, but because it opens a focused program of questions about how geometry, dynamics, and observables are tied together.

\section{Scope and limitations}

It is important to distinguish clearly between what has been established and what remains open.

At the background level, the Weyl-flat statement is classical. The proposition proved in Sec.~II shows that the FRW Weyl tensor vanishes identically and that the far past is asymptotic to the null-boundary power-law solution of Ref.~\cite{Damico:2022}. This does \emph{not} amount to a complete entropy accounting for realistic inhomogeneous quantum states, nor does it derive the thermodynamic arrow of time from a microscopic quantum-gravitational principle. The result is narrower: it removes any background-level obstruction to combining inflation with a Penrose-compatible Weyl-flat origin.

At the level of perturbations, the initial quantum state is imposed rather than derived. The perturbation analysis uses the standard adiabatic vacuum initial conditions deep inside the horizon, which is the technically controlled choice for an early-time adiabatic regime~\cite{BunchDavies:1978yq,Mukhanov:1985rz,Sasaki:1986hm}. What remains open is whether the asymptotic null origin selects that state uniquely, or whether small excitations above the adiabatic vacuum can be accommodated without spoiling the same background structure.

At the level of reheating, the microphysics remains parameterized. The reheating analysis is deliberately agnostic and is encoded in an effective constant $w_{\rm re}$. This is sufficient to determine which inflationary benchmarks can be embedded in a plausible post-inflationary history, but it is not a substitute for specifying couplings of the inflaton to visible or hidden sectors. The background reconstruction and perturbation sector are therefore under control, whereas the particle-physics completion remains to be supplied~\cite{Kofman:1997yn,Pirzada:2025neutralino}.

Beyond exact FRW, stability has not been analyzed. Because the background Weyl tensor vanishes identically, the next question is whether the asymptotically Weyl-flat property is stable under controlled anisotropic or inhomogeneous deformations. This issue is especially nontrivial here because the model contains a finite-$N$ deformation and an automatic exit, so the answer need not coincide with that of exact power-law inflation.

These limitations identify the precise points at which the framework must either mature into a fuller cosmological scenario or meet its sharpest counterarguments. Because the asymptotic mechanism has been written out explicitly, those tests can now be formulated with unusual precision.

\section{Conclusions}

We have presented a single-field realization of the Weyl-flat null origin associated with exact power-law inflation. The central claim is that the geometrically relevant ingredient is not exact power-law evolution throughout the observable era, but the asymptotic condition $\epsilon(N)\to\epsilon_\infty\in(0,1)$ in the far past. Once that condition is isolated, the Weyl-compatible null origin can be preserved while the observable window remains flexible enough to satisfy current data.

The minimal flow family
\begin{equation}
    \epsilon(N)=\epsilon_\infty+(1-\epsilon_\infty)\left(\frac{N_0}{N+N_0}\right)^p,
    \qquad p>1,
\end{equation}
implements this idea transparently. It yields an asymptotic exponential tail in field space, a smooth slow-roll shelf during the observable era, and an automatic end to inflation at $N=0$. We reconstructed the background, derived the perturbation equations directly in e-fold time, computed the scalar and tensor spectra by numerical mode integration, and showed that the viable parameter space forms a corridor rather than an isolated point. We also distinguished between a benchmark that is viable in the CMB plane and one that is additionally compatible with a conventional reheating history.

The resulting picture is conceptually sharp and empirically nontrivial. A Weyl-flat null origin can be realized within a realistic, calculable canonical single-field model. Equally important, the construction clarifies the next layer of questions. Once the asymptotic geometric requirement is separated from the finite-$N$ dynamics, one can examine, case by case, whether state selection, reheating microphysics, anisotropy, or ultraviolet structure preserve that separation or undo it. The value of the framework lies not only in the existence statement itself, but in the more sharply posed question it opens: how much of the observable universe can remain sensitive to the geometry of the beginning without being rigidly fixed by it.
\\{\bf Acknowledgments.} 
Princess Nourah bint Abdulrahman University Researchers Supporting Project number (PNURSP2026R939), Princess Nourah bint Abdulrahman University, Riyadh, Saudi Arabia.

\appendix

\section{Derivation of the exact reconstruction formulas}

Starting from Eqs.~\eqref{eq:Friedmann1} and \eqref{eq:Friedmann2}, one has
\begin{equation}
    \dot\phi^2 = 2\mpl^2H^2\epsilon,
\end{equation}
so that
\begin{equation}
    \frac{d\phi}{dN}=\frac{\dot\phi}{\dot N}=-\frac{\dot\phi}{H}=-\mpl\sqrt{2\epsilon},
\end{equation}
where the negative sign corresponds to the monotonic rolling branch used throughout the paper. Likewise,
\begin{equation}
    \frac{d\ln H}{dN}=\frac{\dot H}{H\dot N}=\frac{-\epsilon H^2}{-H^2}=\epsilon.
\end{equation}
Equation~\eqref{eq:Hanalytic} follows by direct integration of Eq.~\eqref{eq:epsansatz}. Differentiating Eq.~\eqref{eq:Hanalytic} reproduces the original flow ansatz identically.

The asymptotic slope of the potential may be obtained either from the proposition in the main text or directly from the Hamilton-Jacobi form. Since
\begin{equation}
    \frac{d\ln H}{d(\phi/\mpl)}=-\sqrt{\frac{\epsilon}{2}},
\end{equation}
one has
\begin{equation}
    \frac{d\ln V}{d(\phi/\mpl)}
    =2\frac{d\ln H}{d(\phi/\mpl)}+\frac{d\ln(3-\epsilon)}{d(\phi/\mpl)}
    \to -\sqrt{2\epsilon_\infty}
    \qquad (N\to\infty),
\end{equation}
which reproduces Eq.~\eqref{eq:VasymptoticGeneral}. This makes explicit that the exponential tail is not a guess but an unavoidable consequence of asymptotic constant $\epsilon$.

\section{Mode equations in e-fold time}

For completeness we record the derivation of Eq.~\eqref{eq:modeeqN}. Since $d/d\eta=-aH\,d/dN$, one has
\begin{equation}
    u_k'=-aH\,u_{k,N},
\end{equation}
and therefore
\begin{align}
    u_k'' &= -(aH)'u_{k,N}+(aH)^2u_{k,NN}
    \\
    &= (aH)^2\left[u_{k,NN}-(1-\epsilon_1)u_{k,N}\right],
\end{align}
where we used $(aH)'=(aH)^2(1-\epsilon_1)$. Substituting into Eq.~\eqref{eq:MSeta} gives Eq.~\eqref{eq:modeeqN}. Repeating the same steps for $z$ yields
\begin{equation}
    \frac{z''}{z}=(aH)^2\left[y_N+y^2-(1-\epsilon_1)y\right],
    \qquad y\equiv\frac{d\ln z}{dN},
\end{equation}
which is Eq.~\eqref{eq:OmegaDef}. For scalar modes, $z_s=a\sqrt{2\epsilon_1}\mpl$, giving Eqs.~\eqref{eq:ys} and \eqref{eq:Omegas}; for tensor modes, $z_t=a\mpl$ yields Eq.~\eqref{eq:Omegat} immediately.

\section{Reheating formulas}

The derivation of Eqs.~\eqref{eq:Nreformula} and \eqref{eq:Treformula} is standard, but we summarize the main steps to keep the manuscript self-contained. The pivot identity may be written as
\begin{equation}
    \ln\frac{k_*}{a_0H_0}=-N_* - N_{\rm re} + \ln\frac{a_{\rm re}}{a_0}+\ln\frac{H_*}{H_0}.
\end{equation}
Entropy conservation implies
\begin{equation}
    \frac{a_{\rm re}}{a_0}=\left(\frac{43}{11g_{s,\rm re}}\right)^{1/3}\frac{T_0}{T_{\rm re}},
\end{equation}
while the reheating density satisfies Eq.~\eqref{eq:rhoreheat}. Solving Eq.~\eqref{eq:rhorad} for $T_{\rm re}$ and substituting into the pivot relation yields Eq.~\eqref{eq:Nreformula}; reinserting that result into Eq.~\eqref{eq:rhorad} produces Eq.~\eqref{eq:Treformula}. In the present model, the only ingredients required from the inflationary side are $H_*$ and $\rho_{\rm end}$, both of which are fixed by the reconstructed background.

\bibliographystyle{apsrev4-2}
\bibliography{refs}

\end{document}